\documentclass[12pt]{l4dc2021} 

\title[MRAC-RL: On-Line Policy Adaptation]{MRAC-RL: A Framework for On-Line Policy Adaptation Under Parametric Model Uncertainty}
\usepackage{times}
\usepackage{mathtools}
\usepackage{tikz}
\usepackage{wrapfig}
\usepackage[skip=5pt]{caption}

\DeclareMathOperator{\Tr}{Tr}

\usepackage{algorithm}
\usepackage{algorithmic}
\usepackage{graphicx}
\usepackage{cleveref}
\usepackage{booktabs}
\usepackage{natbib}
\usepackage{enumitem} 
\usepackage[export]{adjustbox}
\usepackage[english]{babel}
\usepackage[autostyle, english = american]{csquotes}
\MakeOuterQuote{"}

\newtheorem{manualtheoreminner}{Theorem}
\newenvironment{manualtheorem}[1]{%
  \renewcommand\themanualtheoreminner{#1}%
  \manualtheoreminner
}{\endmanualtheoreminner}

\author{%
 \Name{Anubhav Guha} \Email{anguha@mit.edu}\\
 \Name{Anuradha Annaswamy} \Email{aanna@mit.edu}\\
 \addr Massachusetts Institute of Technology, Cambridge, MA 02139%
}

\begin{document}

\tikzstyle{block} = [draw, fill=white, rectangle, 
    minimum height=3em, minimum width=6em]
\tikzstyle{sum} = [draw, fill=white, circle, node distance=1cm]
\tikzstyle{input} = [coordinate]
\tikzstyle{output} = [coordinate]
\tikzstyle{pinstyle} = [pin edge={to-,thin,black}]

\maketitle

\begin{abstract}%
Reinforcement learning (RL) algorithms have been successfully used to develop control policies for dynamical systems. For many such systems, these policies are trained in a simulated environment. Due to discrepancies between the simulated model and the true system dynamics, RL trained policies often fail to generalize and adapt appropriately when deployed in the real-world environment. Current research in bridging this "sim-to-real" gap has largely focused on improvements in simulation design and on the development of improved and specialized RL algorithms for robust control policy generation. In this paper we apply principles from adaptive control and system identification to develop the model-reference adaptive control \& reinforcement learning (MRAC-RL) framework. We propose a set of novel MRAC algorithms applicable to a broad range of linear and nonlinear systems, and derive the associated control laws. The MRAC-RL framework utilizes an inner-loop adaptive controller that allows a simulation-trained outer-loop policy to adapt and operate effectively in a test environment, even when parametric model uncertainty exists. We demonstrate that the MRAC-RL approach improves upon state-of-the-art RL algorithms in developing control policies that can be applied to systems with modeling errors.%
\end{abstract}


\section{Introduction}\label{sec1}
Reinforcement learning (RL) methods are quickly becoming popular in the development of control policies for complex systems and environments. Successful applications have been broad and varied - ranging from direct actuator-level control and state regulation to high-level planning and decision making (\citealt{mnih2015}, \citealt{Ng2006}, \citealt{lillicrap2019continuous}, \citealt{kober2013}, \citealt{Silver1140}). The effectiveness of reinforcement learning algorithms in overcoming constraints that typically limit classical control techniques has enabled RL's application to decision making and continuous control tasks (\citealt{recht2018tour}, \citealt{schulman2018highdimensional}).

Many RL algorithms are fundamentally data-driven methods. As a result, control polices are often learned largely in simulation. Training in simulation is a powerful technique, allowing for a near infinite number of agent-environment interactions - in comparison, training a policy on an actual plant could be expensive, time-consuming or dangerous. In practice, however, policies trained in simulation often exhibit degenerate performance when applied to real systems (\citealt{Koos2010}) due to modeling errors (\citealt{tan2018simtoreal}). As a result, many researchers have focused on methods to bridge the "sim-to-real" gap.

In this paper we introduce a framework that enables improved performance of RL-trained policies applied to systems with modeling errors. Termed Model-Reference Adaptive Control \& Reinforcement Learning (MRAC-RL), the framework consists of  adaptive control elements in the inner-loop and RL elements in the outer-loop. This inner-outer loop architecture allows a trained policy to adapt control outputs on-line in order to account for model perturbations. The theoretical foundations of MRAC (\citealt{narendra1989}) are leveraged to drive the
“real-world” system states to match the simulated states. The central merit of this MRAC-RL framework
is that it drives the true system to react to the learned control policy in the same way that the simulated system responded during training.
\subsection{Related Work}\label{sec:1.1}
\subsubsection{Reinforcement Learning}\label{sec:1.1.1}
A number of reinforcement learning algorithms have been successfully used to solve continuous control tasks. We pay special attention to the class of deep reinforcement learning (DRL) algorithms which utilize deep neural networks for function approximation. Throughout this paper we specifically reference the Proximal Policy Optimization (PPO), Soft Actor-Critic (SAC) and Deep Deterministic Policy Gradient (DDPG) algorithms (\citealt{schulman2017proximal}, \citealt{haarnoja2018soft}, \citealt{lillicrap2019continuous}). These three DRL algorithms have been applied to a number of varied tasks, and are considered to be state-of-the-art (\citealt{wang2019benchmarking}, \citealt{duan2016benchmarking}, \citealt{henderson2019deep}). While RL/DRL algorithms show significant promise, it is often difficult to reliably predict the behavior of a learned policy - an issue that is exacerbated when the policy is applied to an environment different from the one seen during training (\citealt{fulton2018safe}, \citealt{AurkoMismatch}, \citealt{zhang2018dissection}, \citealt{rajeswaran2018generalization}, \citealt{packer2019assessing}).

Even the most powerful DRL algorithms may fail to generalize in the presence of modeling errors (\citealt{higgins2018darla}, \citealt{nagabandi2019learning}, \citealt{lake2016building}). The research community has largely tackled these challenges by developing specialized RL algorithms. For example, the model-based PILCO (\citealt{Deisenroth2011}) uses a learned probabilistic dynamics model to account for dynamic uncertainty, while DARLA (\citealt{higgins2018darla}) improves sim-to-real transfer by learning robust features. Another popular approach is to directly modify the simulation \& training protocols. In \citealt{rajeswaran2017epopt} an ensemble of environments with varying dynamics were used to improve the robustness of learned policies, while \citealt{loquercio2020drone} used simulated domain randomization to bridge the sim-to-real gap on a drone racing task. Many approaches utilize a combination of these techniques. In \citealt{tan2018simtoreal} a robust RL algorithm was used with a system identification technique to enable real-world quadruped control. In \citealt{nagabandi2019learning} meta-learning principles were used to modify the policy training process and subsequently adapt the policy to unmodeled errors at test-time.

As detailed above, the majority of the research in bridging the sim-to-real gap has focused on improved simulation techniques and improved RL algorithms (\citealt{pinto2017robust}, \citealt{packer2019assessing}, \citealt{higgins2018darla}, \citealt{Deisenroth2011}, \citealt{nagabandi2019learning}, \citealt{rajeswaran2018generalization}, \citealt{berkenkamp2017safe}). There has, however, been little attention paid to methods that may be used to inject additional robustness and adaptability into an already-trained policy.
\subsubsection{Adaptive Control and System Identification}
Adaptive control and system identification methods have long been used in the control of safety and performance sensitive systems (\citealt{L1AC}, \citealt{dydek2013}, \citealt{WISE2011}, \citealt{michini2009}, \citealt{Wiese2013}). Unlike many RL algorithms, adaptive control techniques excel in the "zero-shot" enforcement of control objectives - that is, in learning to accomplish a task on-line (\citealt{recht2018tour}, \citealt{narendra1989}). These adaptive techniques are able to accommodate, in real-time, constraints on the control input magnitude (\citealt{Karason1994}, \citealt{Lavretsky2004}) and rate (\citealt{gaudio2019adaptive}). This ability to achieve control goals while accounting for parametric uncertainties in real-time is the strength of adaptive control. A weakness of adaptive methods is the general inability to integrate complex optimization objectives, as the underlying methods often focus on the minimization of tracking and regulation errors (\citealt{slotine1991applied}).
In contrast, RL-trained policies can handle a broad range of tasks \& objectives (\citealt{sutton}), but often fail to generalize appropriately in the presence of modeling errors (as discussed in Section \ref{sec:1.1.1}). In this paper we propose a method to combine the strengths of RL and adaptive control, while minimizing the weaknesses. Specifically, we make prolific use of model-reference adaptive control (MRAC) techniques. In the MRAC paradigm, a known reference model (characterized by known model parameters and a known model form) defines the desired closed-loop behavior of the system. The "true" model is then driven to match the reference system by the MRAC algorithm. In classical application of MRAC, the form and structure of the reference model are treated as design parameters (\citealt{Krstic1995}). In this paper, however, we treat the reference model as the closed-loop system formed by the simulation model and the RL-derived control policy. The MRAC task is then to drive the "true" system (which is seen only at test-time, and not during training) to track this closed-loop reference model. By synthesizing such an MRAC-RL architecture, we use guidelines from RL to generate a policy for a specified reference/simulated model, and guidelines from adaptive control to adapt this policy in real-time in order to account for "sim-to-real" modeling discrepancies. The RL component may be viewed as an outer-loop block, while the MRAC component may be viewed as an inner-loop block. 

The general problem of interest is posed in Section \ref{sec:2} with a motivating example. The MRAC foundation of the proposed MRAC-RL framework is laid in Section \ref{sec:3}. An algorithm that implements this framework is provided in Section \ref{sec:4}, and it is shown in Section \ref{sec:5} that MRAC-RL results in improved performance for an inverted pendulum task. Summary and conclusions are presented in Section \ref{sec: 6}.

\section{Problem Statement}\label{sec:2}
Consider a continuous-time (CT), deterministic dynamical system defined by the map $f: X \times U \to X$:
\begin{equation}\label{eq:dynamics}
    \dot{x} = f(x(t), u(t), \phi), \quad \quad x(0) = x_0, \quad \quad u(t) \in U
\end{equation}
where $\phi$ corresponds to system parameters that may be subject to uncertainties. Associated with this system is some cost functional $c: X \times U \times \mathbb{N} \to \mathbb{R}$ so that the optimal (finite time-horizon) control problem is given by:

\begin{equation}\label{eq:optimization}
\begin{aligned}
\min_{{\scriptscriptstyle u(t) \in U \; \forall t \in [0, T]}} \quad & \int_{0}^{T}c(x(t), u(t), t)dt\\
\textrm{subject to} \quad & \dot{x} = f(x(t), u(t), \phi) \quad \forall t \in [0, T]\\
  &x(0) = x_0   \\
\end{aligned}
\end{equation}
Suppose reinforcement learning techniques are used to generate a control policy $\pi$ such that $u(t) = \pi(x(t))$ produces approximately optimal solutions to the system in \eqref{eq:optimization}. If the system to be controlled is a physical system, we will likely train the policy largely in simulation. In developing this simulation, we implicitly make a choice of an assumed state equation, henceforth referred to as the \textit{reference model}: $\dot{x}_r = f_r(x_r(t), u_r(t), \phi_r)$, where $\phi_r$ denotes the nominal values of the system parameters $\phi$. The subscript $r$ denotes the fact that these quantities are simulated and their relationships are determined by the (known) reference model. Applying the reinforcement learning method of choice to the discrete-time (DT) variant of the optimization in \eqref{eq:optimization} results in the approximate optimal control policy: $\pi(x)$. Note that most RL approaches will formulate \eqref{eq:dynamics}-\eqref{eq:optimization} as a DT Markov decision process (MDP) (\citealt{sutton1998introduction}, \citealt{kaelbling1996reinforcement}). For the remainder of this paper, we utilize CT notation with the assumption that the policy-generated action is applied continuously over the MDP discrete time interval, and that the numerical integration frequencies are large enough to consider digital implementations of CT algorithms. 

In the standard RL approach, the trained policy $\pi$ is then applied to the \textit{true model}: $\dot{x}(t) = f(x(t), \pi (x(t)), \phi)$ (\citealt{kober2013}). Recall that $\pi$ was trained entirely using the \textit{reference model}. If the reference model is erroneous (e.g, system parameters were modeled imperfectly), then $f_r(x,u, \phi_r) \neq f(x,u, \phi)$ and the reference and true trajectories will likely diverge and performance may degrade. The goal of this paper is to determine the control policy $u$ in \eqref{eq:optimization} despite uncertainties in $\phi$.

\subsection{A Motivating Example}\label{sec:2.1}
We introduce a variant of the canonical swing-up inverted pendulum task (\citealt{furuta1992swing}), henceforth referred to as the set-point randomized inverted pendulum (SRIP). We will use this example to illustrate the advantages of the MRAC-RL architecture. In the classic swing-up problem, a rigid rod is fixed at one end by a joint. The goal is to apply torque at the joint so that the free end of the rod swings upright and subsequently holds the unstable equilibrium. The SRIP objective is to instead drive the pendulum angle to a random set-point. This random set-point is provided in an augmented state vector, and changes at a set rate. As a benchmark task for control under model uncertainty, SRIP is preferable to the swing-up task. In the swing-up task the goal/cost-minimizing state ($\theta = 0$, $\dot{\theta} = 0)$ represents an equilibrium of the system. Even though the equilibrium is unstable, the ideal control magnitude tends to zero as the equilibrium is approached. In contrast, optimal control of the SRIP requires a non-zero steady-state control signal. Consider the following linear and nonlinear models of the inverted pendulum:
\begin{equation}\label{linnonlin}
    \begin{aligned}
         ml^2\ddot{\theta} = mgl\theta -b\dot{\theta} + u\quad
        \textrm{(linear)} \quad \quad ml^2\ddot{\theta} = mgl\sin\theta -b\dot{\theta} + u \quad \textrm{(nonlinear)}
    \end{aligned}
\end{equation}

\noindent where $m,$ $g,$ $l,$ $k > 0$ are the mass, gravitational, length and viscous drag constants respectively. The goal of the task is to maintain a non-zero set-point $[\theta_0, 0]^T$. In order to hold this set-point, the required control effort is necessarily a function of the model parameters. As a result, the SRIP task is more punishing than the base swing-up problem when the true model parameters deviate from the simulated (reference) model parameters, and thus serves as a suitable benchmark for robust \& adaptive reinforcement learning. Note that the linear \& nonlinear variants of the SRIP task represent specific examples of the generic optimal control problem posed in \eqref{eq:optimization}. Here, the state $x = [\theta, \dot{\theta}]^T$ and the cost $c$ is a function that penalizes deviation from the set-point (e.g, $c = q_1(\theta - \theta_0)^2 + q_2\dot{\theta}^2 + ru^2$ with $q_1, \;q_2, \; r > 0$).

One can use a reference model of the inverted pendulum to train a control policy $\pi$ for the SRIP task via reinforcement learning (\citealt{lillicrap2019continuous}). Suppose that this policy $\pi$ is then used to solve the SRIP task in a test environment in which the true dynamics model deviates from the reference model. For example, the true mass $m$ and length $l$ of the test inverted pendulum may differ from the mass and length of the pendulum on which $\pi$ was trained. In the MRAC-RL framework that we propose, the learned policy is never applied directly to the true system. Instead, we utilize an inner-outer loop architecture whereby the control policy is used to generate a closed-loop reference system. At runtime, adaptive control methods are used in the inner-loop to drive the true system to track the closed-loop reference system. This approach ensures that the reinforcement learning agent is only ever interacting with the environment in which it was trained, while the adaptive control loop independently handles the issue of parametric model uncertainty. Note that this is a strict departure from the standard RL paradigm, in which a policy trained in a simulated environment is directly used as a feedback controller in the true environment.

Central to the MRAC-RL framework is the ability to guarantee convergence of the true model to the closed-loop reference model. We hypothesize that the ability to track a simulated reference model will improve the performance and reliability of an RL-trained policy. In the next section we develop the MRAC algorithms necessary to construct the MRAC-RL framework.
\section{Model Reference Adaptive Control}\label{sec:3}
We now present the MRAC control approach for linear \& nonlinear dynamic systems in the presence of parametric model uncertainties.

\subsection{Linear Model}\label{sec:linmod}
We develop an MRAC algorithm for a class of $n$-dimensional linear dynamic models of the form:

\begin{equation}\label{eq:lin_ref_mod_cc}
\begin{aligned}
\quad & \dot{x} = Ax + Bu\\
\textrm{with} \quad & x \coloneqq \begin{bmatrix} x_1 \\ \vdots\\ x_n \end{bmatrix}
A \coloneqq \begin{bmatrix}
        0 & 1 & 0 & \dots & 0\\
        \vdots & \vdots & \vdots & \ddots & \vdots\\
        a_1 & a_2 & a_3 & \dots & a_n
        \end{bmatrix}
B \coloneqq \begin{bmatrix} 0 \\ \vdots \\ b \end{bmatrix}\\
\end{aligned}
\end{equation}
\noindent All $a_i, \; b$ are non-zero and have known signs but unknown values. This system is henceforth referred to as the true system, and the goal is to choose the control input $u$ so as to accomplish a control objective. It is easy to see that the SRIP task with the linear model is a specific case of \eqref{eq:lin_ref_mod_cc}. A known and potentially inaccurate reference model of the system is given by $\dot{x}_r = A_rx + B_ru_r$. The subscript $r$ denotes parameters and signals belonging to the reference model. $(A_r, B_r)$ are in the same controllable form as $(A, B)$ in \eqref{eq:lin_ref_mod_cc}. Let $\alpha_r$ refer to the last row of the matrix $A_r$, and $b_r$ be the non-zero element of $B_r$. Note that $\alpha_r, b_r$ are known, and therefore one can adopt a host of control methods or an RL approach to determine $u_r$ so that $x_r$ behaves in a desired manner. The true system to be controlled \eqref{eq:lin_ref_mod_cc} may be equivalently rewritten as $\dot{x} = Ax + \lambda B_r u$, where we have introduced an unknown scalar $\lambda > 0$. The MRAC goal is to determine the input $u(t)$ so that the tracking error converges to zero: $\lim\limits_{t \to \infty}||e(t)|| = 0$, with $e(t) \coloneqq x(t) - x_r(t)$. 

As mentioned in Section \ref{sec:1.1}, the goal of MRAC is to drive the current tracking errors to zero, rather than the global goal in \eqref{eq:optimization} of optimizing a function over the entire trajectory. Because the MRAC solution is expected to occur in real-time, it is difficult to deliver globally optimal solutions while simultaneously learning about an uncertain system. 

\noindent We pick a diagonal matrix $D \in \mathbb{R}^{nxn}$, with diagonal entries defined by:
    \begin{equation*}
        D_{ii} = \begin{cases} 
          \omega_{i} & \alpha_{i, r} > 0 \quad\\
          \psi_{i} & \alpha_{i, r} < 0\\
          \end{cases}
    \end{equation*}

\noindent $a_{i, r}$ refers to the $i$th component of the known vector $\alpha_r$. $\omega_i, \psi_i$ are picked so that $\omega_i > 1$, $\psi_i \leq 0$ for $i = 1, 2, \ldots, n$. For convenience we define the vector $h \coloneqq [0, 0, \ldots, 1]^T \in \mathbb{R}^n$. Note then that the matrix $A_H \coloneqq A_r - h(D\alpha_r)^T$ is Hurwitz. We additionally define $\xi \coloneqq u_r - \frac{1}{b_r}(D\alpha_r)^Te$. We now introduce the MRAC control law and the associated adaptive parameter update laws:
    \begin{equation}\label{eq:lin_cntrl_law}
        \begin{aligned}
            u = \hat{K}_x^Tx + \hat{k}_u\xi
    \end{aligned}
    \end{equation}
    \begin{equation}\label{eq:lin_param_update}
        \begin{aligned}
        \dot{\hat{K}}_x = -\Gamma_xxe^TPB_r, \quad \dot{\hat{k}}_u = -\gamma_u\xi e^TPB_r
    \end{aligned}
    \end{equation}
\noindent where $\Gamma_x = \Gamma_x^T \succ 0$, $\gamma_u > 0$ and $P = P^T \succ 0$ solves the Lyapunov equation: $PA_H + A_H^TP = -Q$ with $Q = Q^T \succ 0$. $\hat{K}_x$ and $\hat{k}_u$ may be initialized arbitrarily - however for this work we propose setting $\hat{K}_x(0) = \textbf{0}_{n\times 1}$, $\hat{k}_u(0) = 1$. If no model discrepancy exists, these initial parameter values immediately lead to perfect tracking of $x_r$ by $x$. We now state the two main properties of MRAC that are relevant for our proposed MRAC-RL architecture (with parameter estimation errors $\Tilde{K}_x, \Tilde{k}_u$):
\begin{theorem}\label{theorem:control_law}
    For the system \eqref{eq:lin_ref_mod_cc}, associated reference system $\dot{x}_r = A_r x_r + B_r u_r$, and adaptive laws \eqref{eq:lin_cntrl_law}-\eqref{eq:lin_param_update}, the function $V(e, \Tilde{K}_x, \tilde{k}_u) = e^TPe + \lambda \Tr(\Tilde{K}^T_x\Gamma_x^{-1}\Tilde{K}_x) + \lambda\frac{\Tilde{k}_u^2}{\gamma_u}$ is a valid Lyapunov function.
\end{theorem}

\begin{theorem}\label{corollary:control_law}
    Theorem \ref{theorem:control_law} and \eqref{eq:lin_cntrl_law}-\eqref{eq:lin_param_update} guarantee that $\lim\limits_{t \to \infty}||x(t) - x_r(t)|| = 0$ if $||x_r(t)|| < M_x$ and $||u_r(t)|| < M_u \; \forall t \in [0, T]\;$ for some $M_x, M_u > 0$
\end{theorem}

\subsection{Nonlinear Model}\label{ss: nonlinear model}
The MRAC approach can be extended to a class of $n$-dimensional nonlinear models in a straightforward manner. This class is given by:
\begin{equation}\label{eq:nlin_ref_mod_cc}
    \begin{aligned}
    \dot{x} = A\zeta (x) + \lambda B_r u
    \end{aligned}
\end{equation}
Where $\zeta : \mathbb{R}^n \to \mathbb{R}^n$ is a known nonlinear map of the form:
\begin{equation*}
    \begin{aligned}
    \zeta (x) = [\phi(x_1), x_2, x_3, \ldots, x_n]^T
    \end{aligned}
\end{equation*}
With $\phi: \mathbb{R} \to \mathbb{R}$ a known nonlinear function. For notational simplicity, we use $\zeta$ to refer to $\zeta (x)$. The pair $(A,B_r)$ are in the same form as \eqref{eq:lin_ref_mod_cc}. As in Section \ref{sec:linmod}, $\lambda > 0$ is an unknown scalar, $A$ is unknown and $B_r$ is a known matrix in the desired reference model: $\dot{x}_r = A_r \zeta + B_r u_r$. As in Section \ref{sec:linmod} the goal is to determine a $u$ such that the tracking error converges to zero: $\lim\limits_{t \to \infty}||e(t)|| = 0$, with $e(t) \coloneqq x(t) - x_r(t)$. We pick an $n$-dimensional vector $\beta_r$ with strictly negative components. Additionally, let $\alpha_r$ be the known vector corresponding to the last row of the matrix $A_r$. For convenience, define the vector $h \coloneqq [0, 0, \ldots, 1]^T \in \mathbb{R}^n$. We then define the matrix $A_H \coloneqq A_r - h\alpha_r^T + h\beta_r^T$, which is Hurwitz by construction. Defining $\xi \coloneqq u_r - \frac{1}{b_r}\alpha_r^T(\zeta - \zeta_r) + \frac{1}{b_r}\beta_r^Te$, we introduce the MRAC adaptive laws:
    \begin{equation}\label{eq:nlin_cntrl_law}
        \begin{aligned}
            u = \hat{K}_\zeta^T\zeta + \hat{k}_u\xi
    \end{aligned}
    \end{equation}
    \begin{equation}\label{eq:nlin_param_update}
        \begin{aligned}
        \dot{\hat{K}}_\zeta = -\Gamma_\zeta \zeta e^TPB_r, \quad \dot{\hat{k}}_u = -\gamma_u\xi e^TPB_r
    \end{aligned}
    \end{equation}
\noindent where $\Gamma_\zeta = \Gamma_\zeta^T \succ 0$, $\gamma_u > 0$ and $P = P^T \succ 0$ solves the Lyapunov equation: $PA_H + A_H^TP = -Q$ with $Q = Q^T \succ 0$.
\begin{theorem}\label{theorem:control_law_nl}
    For the MRAC system described in Section \ref{ss: nonlinear model}, the function $V(e, \Tilde{K}_\zeta, \tilde{k}_u) = e^TPe + \lambda \Tr(\Tilde{K}^T_\zeta\Gamma_\zeta^{-1}\Tilde{K}_\zeta) + \lambda\frac{\Tilde{k}_u^2}{\gamma_u}$ is a valid Lyapunov function.
\end{theorem}

\begin{theorem}\label{corollary:nl_control_law} Theorem \ref{theorem:control_law_nl} and \eqref{eq:nlin_cntrl_law}-\eqref{eq:nlin_param_update} guarantee that $\lim\limits_{t \to \infty}||x(t) - x_r(t)|| = 0$ if $||\zeta(t)|| < M_\zeta$ and $||u_r(t)|| < M_u \; \forall t \in [0, T]\;$ for some $M_\zeta, M_u > 0$
\end{theorem}
\section{MRAC-RL}\label{sec:4}
\begin{wrapfigure}{r}{0.4\textwidth}
    \vspace{-10pt}
    \begin{minipage}[H]{.4\textwidth}
        
        \includegraphics[width=1.0\textwidth]{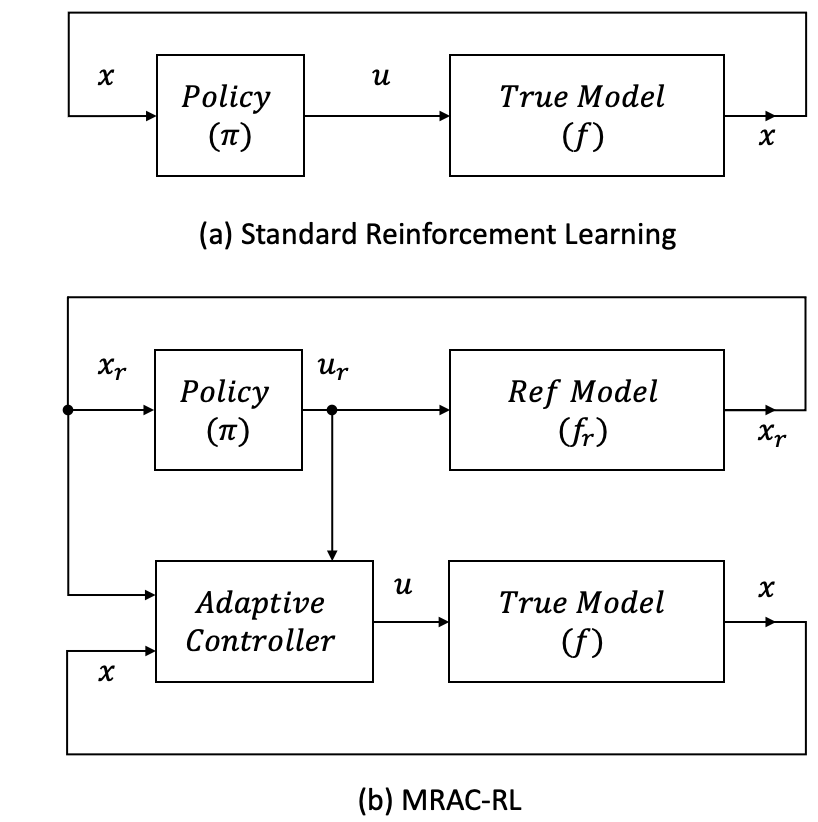}
        \caption{MRAC-RL \& RL}
        \label{fig1}
    \vspace{-10pt}
    \end{minipage}
\end{wrapfigure}
We now present the MRAC-RL framework, as shown in Figure \ref{fig1}. The standard RL use case is shown in Figure \ref{fig1}a: A trained policy directly maps system states ($x$) to control actions $(u)$ in order to control a physical system. The MRAC-RL approach is outlined in Figure \ref{fig1}b: The trained policy operates in a simulated reference system, mapping reference states ($x_r$) to reference actions ($u_r$). An inner loop adaptive controller modifies $u_r$ to produce a control signal ($u$) that drives the true system to track the reference trajectory. As a result, the trained policy never interacts with the true system, instead relying on the adaptive control block to appropriately adjust and modify $u$. Theorems \ref{theorem:control_law}-\ref{corollary:nl_control_law} are leveraged to guarantee satisfactory behavior in the presence of parametric modeling errors for dynamic systems in the form of \eqref{eq:lin_ref_mod_cc} or \eqref{eq:nlin_ref_mod_cc}.

\begin{wrapfigure}{l}{0.5\textwidth}
\begin{minipage}[H]{.5\textwidth}
\vspace{-30pt}
\begin{algorithm}[H]
    \caption{MRAC-RL for the linear SRIP Task}
    \begin{algorithmic}[1]\label{algorithm: al1}
    \STATE {\bfseries Input:} $\pi$; {\bfseries Initialize:} $\hat{K}_x(0)$, $\hat{k}_u(0)$, $x(0)$
    \WHILE{\NOT done}
    \STATE $u_r = \pi(x_r)$,
    \FOR{$i = 1,\ldots, F_1$}
    \STATE {\bfseries Receive:} $x$
    \STATE $e_\theta$, $e_\omega$ $= x - x_r$
    
    \STATE $u = \hat{K}^T_xx + \hat{k}_uu_r - \frac{2}{b_r}\hat{k}_ua_{1, r}e_\theta$
    \STATE $\hat{K}_x \leftarrow \hat{K}_x - \Delta_1\Gamma_xxe^TPB_r$
    \STATE $\hat{k}_u \leftarrow \hat{k}_u - \Delta_1\gamma_u (u_r - \frac{2}{b_r}a_{1,r}e_\theta) e^TPB_r$
    \FOR{$j = 1, \ldots, F_2$}
    \STATE $x_r \leftarrow x_r + \Delta_2(A_rx_r + B_ru_r)$
    \ENDFOR
    \ENDFOR
    \ENDWHILE
\end{algorithmic}
\end{algorithm}
\end{minipage}
\vspace{-27pt}
\end{wrapfigure}
\vspace{10pt}
\noindent As an example, the MRAC-RL framework as applied to the linear SRIP task is detailed in Algorithm \ref{algorithm: al1}. The state $x = [\theta_r, \dot{\theta}_r]^T$ and the matrices:
\begin{equation*}
    A_r = 
        \begin{bmatrix} 
            0 & 1\\
            \frac{g}{l_r} & -\frac{b_r}{m_rl_r^2}\\
        \end{bmatrix} \quad B_r = 
        \begin{bmatrix}
            0\\
            \frac{1}{m_rl_r^2}\\
        \end{bmatrix}
\end{equation*}
define the reference model (with $g, l_r, m_r, b_r$ known). Note, $a_{1,r} = g/l_r, b_r = 1/(m_rl_r^2)$. The true system is assumed to be well-modeled by dynamics of the same form, but with potential parameter differences. Additionally, $F_1$ represents the operating rate of the adaptive inner loop relative to the policy evaluation rate, $F_2$ represents the integration rate of the reference system relative to the adaptive loop rate, and $\Delta_{1,2}$ are the corresponding intervals of numerical integration. To apply this algorithm to the nonlinear SRIP objective, we make the appropriate modifications in steps 7-9 of the algorithm, using \eqref{eq:nlin_cntrl_law} and \eqref{eq:nlin_param_update} in Section \ref{ss: nonlinear model}.

We posit that the proposed combination of the MRAC and RL components as shown in Figure \ref{fig1} is able to ensure that the true system behavior emulates the simulated behavior that the policy $\pi$ was trained to control. In particular, our claim is that the MRAC-RL approach maintains the effectiveness of RL algorithms in generating control policies in the presence of modeling errors by combining the adaptive control components with an RL-trained policy. In the next section we validate this claim for the motivating example presented in Section \ref{sec:2.1}.

\section{Experimental Results}\label{sec:5}
We test the MRAC-RL approach in solving the SRIP task for both the linear and nonlinear models of the inverted pendulum system given in \eqref{linnonlin} and evaluate the efficacy of the framework using three popular reinforcement learning algorithms: PPO, SAC and DDPG (\citealt{schulman2017proximal}, \citealt{haarnoja2018soft}, \citealt{lillicrap2019continuous}). We utilize the Stable Baselines (\citealt{stable-baselines}) implementations of these algorithms. Stable Baselines provides a number of high quality RL algorithms, and is based on the popular OpenAI Baselines implementations. 

The RL algorithms were used to train control policies for the (linear and nonlinear) SRIP reference environment, with $m_r, l_r, b_r = 1,$  $g = 10$. A quadratic cost functional was used for training: $c(\theta, \dot{\theta}, u) = q_1 (\theta - \theta_0)^2 + q_2 \dot{\theta}^2 + r u^2$, for $q_1, q_2, r > 0$. The policies were trained using an agent-environment interaction frequency of $10Hz$. Test environments were then generated using perturbed model parameters, picked from the following ranges: $l, m \in [.75, 1.25]$, $b \in [.001, 2.0]$. Further training and simulation details are provided in Appendix \ref{app:b}. Four frameworks/algorithms were tested:
\begin{itemize}
    \item \textbf{\textit{100Hz RL:}} $\pi(x(t))$ is evaluated at 100Hz and the result is sent to the true model at 100Hz. This is a standard application of a trained policy. No adaptive control occurs at any level.
    \item \textbf{\textit{10Hz RL; 100Hz MRAC:}} $u_r(t) = \pi(x_r(t))$ is evaluated at 10Hz. The MRAC inner loop converts $u_r \to u$ at 100Hz, which is sent to the true model. In the context of Algorithm \ref{algorithm: al1}, this corresponds to a \textbf{do} loop rate of 10Hz, with $F_1 = 10$, and $\Delta_1 = .01s$. $F_2, \Delta_2$ are dependent on the numerical integration specifications
    \item \textbf{\textit{10Hz RL:}} Similar to \textbf{\textit{100Hz RL}} except actions are calculated and sent at 10Hz
    \item \textbf{\textit{10Hz RL; 10Hz MRAC:}} Similar to \textbf{\textit{10Hz RL; 100Hz MRAC}} except the inner loop occurs at 10Hz. That is, the MRAC loop operates in lock-step with the outer loop, providing only a single adaptive update per policy evaluation. In the context of Algorithm \ref{algorithm: al1}, this corresponds to a \textbf{do} loop rate of 10Hz, with $F_1 = 1$, and $\Delta_1 = .1s$.
\end{itemize}
For the linear SRIP task, we additionally test an LQR-based outer-loop control policy: $\pi(x) = -K_rx + u_{0, r}(x_0)$. The LQR feedback gain $K_r$ is determined using the reference model parameters, $x_0$ represents the commanded set-point, and $u_{0, r}$ is the steady-state control required to hold $x_0$ (determined using the reference model). The use of three distinct RL algorithms along with an LQR-derived policy demonstrate the flexible and general nature of the MRAC-RL framework. We need only provide some map $\pi: X \to U$ at the outer loop, and the MRAC component will effectively account for the model discrepancies.
\begin{table}[h!]
  \begin{center}
    \begin{tabular}{c c |c  c|c  c}
      \toprule 
      \multicolumn{2}{c|}{\textbf{Algorithm}} & \multicolumn{2}{c|}{\textbf{Linear Model}} & \multicolumn{2}{c}{\textbf{Nonlinear Model}}\\
      \toprule
      \textbf{Outer Loop} & \textbf{Inner Loop} & \textbf{Average Cost} & 
      \textbf{Average $\mathbf{e_\theta^2}$} & \textbf{Average Cost} & 
      \textbf{Average $\mathbf{e_\theta^2}$}\\

      \midrule 
      \textit{100Hz RL} & \hfill$--$ &            $\mathbf{134}$ &         $78$ &                       $317$     & $1.3\mathrm{e}3$\\
      \textit{10Hz RL} & \textit{100Hz MRAC} & $        140$ & $\mathbf{1.67}$ &                      $\mathbf{220}$  & $\mathbf{98}$\\
      \textit{10Hz RL}&    \hfill$--$ &           $        264$ &         $331$ &                       $322$&         $1.4\mathrm{e}3$\\
      \textit{10Hz RL} & \textit{10Hz MRAC}&   $        149$ &         $28$ &                       $229$&         $131$\\
      \textit{100Hz LQR}&   \hfill$--$ &          $        250$ &        $1.2\mathrm{e}4$ &            $--$ & $--$\\
      \textit{10Hz LQR} & \textit{100Hz MRAC}& $        228$ &        $14$ &                      $--$ & $--$\\
      \bottomrule 
    \end{tabular}
      \caption{Results from the SRIP task with model discrepancy. A number of algorithms with varying inner/outer loop structures are tested. We opt to draw comparisons between algorithms that update the true control ($u$) at the same rate. For example, we compare \textit{[100Hz Rl; $--$]} and \textit{[10Hz Rl; 100Hz MRAC]}. Average cost is calculated as the average $c(\theta, \dot{\theta}, u)$ accumulated over all test sets. Average $e_\theta^2$ (units are deg$/s^2$) is calculated as the average reference model $\theta$ tracking error (e.g, as $\overline{(\theta - \theta_r)^2}$). For both performance metrics, lower values are preferable.
      \label{tab:table1}
  } 
  \end{center}
\vspace{-17pt}
\end{table}
\noindent Reinforcement learning algorithms are generally rated on their ability to maximize accumulated reward (or to minimize cost). Though this is certainly an important metric, we pay special attention to the reference tracking ability of a given algorithm in the presence of modeling errors. We claim that minimizing this divergence is important in the development of RL-based control algorithms that can effectively bridge the sim-to-real gap. Upon inspection of the \textbf{Average $\mathbf{e_\theta^2}$} columns in Table \ref{tab:table1}, we see that the insertion of an MRAC inner loop improves reference tracking performance. Moreover, in \textit{most} cases, the average cost $c(\theta, \dot{\theta}, u)$ incurred is substantially lowered by the use of an MRAC inner-loop. That is, the MRAC-RL framework demonstrates noticeably improved performance on the SRIP task, while significantly improving reference model tracking ability. Additionally, these results are \textit{robust} over a broad range of perturbed model parameter values ($\pm 25\%$ error in $l$, $m$, and $\pm 100\%$ error in $b$), initial conditions, and reinforcement learning architectures (Appendix \ref{app:b}).

\begin{figure}[H]
    \centering
    \includegraphics[width=1.0\textwidth]{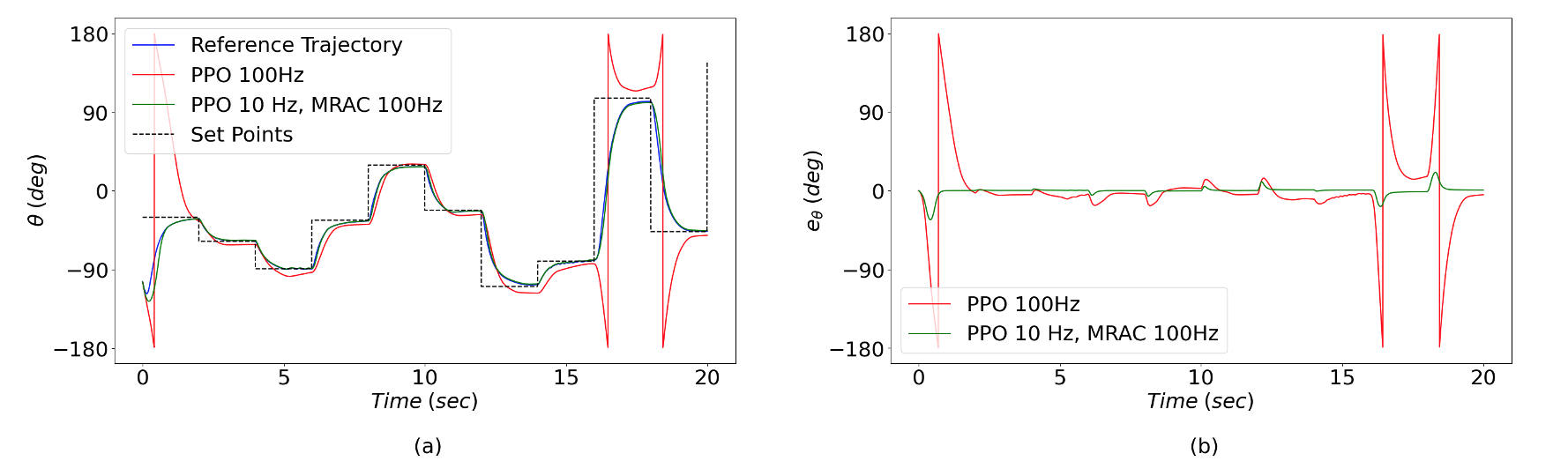}

    \caption{Reference model tracking performance. The reference trajectory is generated via application of a PPO-trained control policy to the reference model. A perturbed "true" model is produced, and we compare direct policy application (red), and the MRAC-RL inner-outer loop framework (blue). (a) depicts the $\theta$ trajectories and (b) shows the reference $\theta$ tracking error $e_\theta$.}
    \label{fig:simulation}
\end{figure}
\vspace{-25pt}
\section{Summary \& Conclusions}\label{sec: 6}
The overall goal of this effort is to solve optimal control problems in the form of \eqref{eq:optimization}, when system parameter ($\phi$) modeling errors are present. We propose the MRAC-RL framework as a solution for specific classes of \eqref{eq:dynamics}, in the forms of \eqref{eq:lin_ref_mod_cc} and \eqref{eq:nlin_ref_mod_cc}. We articulate the stability guarantees for these systems in Theorems \ref{theorem:control_law}-\ref{corollary:nl_control_law}, and demonstrate that, under mild conditions, the tracking objective $\lim\limits_{t \to \infty}||e(t)|| = 0$ is achieved in the presence of parametric uncertainties. The MRAC algorithms proposed in \eqref{eq:lin_cntrl_law}, \eqref{eq:lin_param_update}, \eqref{eq:nlin_cntrl_law}, and \eqref{eq:nlin_param_update} are used to construct the inner-loop of the MRAC-RL framework. We then rely on extensive results and research in reinforcement learning (\citealt{mnih2015}, \citealt{lillicrap2019continuous}, \citealt{haarnoja2018soft}, \citealt{schulman2017proximal}) to produce a pseudo-optimal controller at the MRAC-RL outer-loop. We posit that this combined RL \& adaptive control architecture enables predictable and performant solutions to \eqref{eq:optimization}. The MRAC algorithms proposed in \eqref{eq:lin_cntrl_law} - \eqref{eq:nlin_param_update} were used to construct and successfully apply the MRAC-RL solution to the linear and nonlinear variants of the motivating problem, with an example implementation given in Algorithm \ref{algorithm: al1}.

An inverted pendulum task was introduced and used to benchmark the MRAC-RL framework against a number of popular RL algorithms. We demonstrated that, on this task, the MRAC-RL approach augmented and improved upon three reinforcement learning algorithms: PPO, SAC and DDPG. The MRAC inner loop was able to confer enhanced adaptive properties upon RL-trained policies without requiring any domain randomization or retraining. This is in contrast with the majority of the methods discussed in Section \ref{sec:1.1.1}, in which adaptive and robust properties are introduced via simulator \& RL algorithm design. In theory, the MRAC-RL framework could operate in a modular manner with such methods and algorithms - for example, a robust RL algorithm such as PILCO (\citealt{Deisenroth2011}) could be used to train the outer loop control policy.

In this paper we have paid special attention to the minimization and convergence of \textit{tracking} error, but did not address the convergence of \textit{parameter} error. This is largely due to the inner-loop structure of the framework, which gives the MRAC algorithm no authority in determining the reference input. As a result, the persistent excitation (PE) condition, which is necessary \& sufficient for parameter convergence, cannot be ensured. An interesting line of future research could be in investigating methods in which the policy is trained to promote the PE condition, so that an adaptive loop may effectively learn the model parameters. Such an approach was used for robust linear-quadratic regulation in \citealt{dean2019safely}.  

The next step in this line of research is to evaluate the MRAC-RL framework over a broader set of tasks. While the inverted pendulum is a good canonical control benchmark, the algorithms discussed in this paper can be extended to systems with more dimensions, greater degrees of nonlinearity, and non-trivial dynamic interactions (e.g, contact forces).
\acks{This work was supported by the Boeing Strategic University Initiative}
\clearpage
\bibliography{references}

\clearpage
\appendix
\noindent{\LARGE \textbf{Appendix}}
\section{Convergence and Stability Proofs}
\subsection{Adaptive Controller: Linear System}\label{ss: linear}

\begin{manualtheorem}{\ref{theorem:control_law} (Summary)}
    Let $\alpha_r$ be the vector corresponding to the last row of the known matrix $A_r$. Furthermore, $b_r$ is the known non-zero element of $B_r$. Pick a diagonal matrix $D \in \mathbb{R}^{nxn}$, with components defined as:
    \begin{equation*}
        D_{ij} = \begin{cases} 
          \omega_{i} & \alpha_{i, r} > 0 \quad \textrm{and} \quad i = j\\
          \psi_{i} & \alpha_{i, r} < 0 \quad \textrm{and} \quad i = j\\
          0 & i\neq j\\
          \end{cases}
    \end{equation*}
    with $\omega_i > 1$, $\psi_i \leq 0$ for $i = 1, 2, \ldots, n$. We additionally define $\xi = u_r - \frac{1}{b_r}(D\alpha_r)^Te$. Using the following control and adaptive parameter update laws:
    \begin{equation}
        u = \hat{K}_x^Tx + \hat{k}_u\xi\
    \end{equation}
    \begin{equation}
        \dot{\hat{K}}_x = -\Gamma_xxe^TPB_r, \quad \dot{\hat{k}}_u = -\gamma_u\xi e^TPB_r
    \end{equation} 
$V(e, \Tilde{K}_x, \tilde{k}_u) = e^TPe + \lambda \Tr(\Tilde{K}^T_x\Gamma_x^{-1}\Tilde{K}_x) + \lambda\frac{\Tilde{k}_u^2}{\gamma_u}$ is a Lyapunov function
\end{manualtheorem}

\begin{proof} We consider dynamical systems given by the following linear model:
\begin{equation}
    \begin{aligned}
    \dot{x} = Ax + Bu
    \end{aligned}
\end{equation}
with $A$, $B$ in the controllable canonical form:
\begin{equation}\label{eq:A.1.1}
    \begin{aligned}
        A = \begin{bmatrix}
        0 & 1 & 0 & \dots & 0\\
        0 & 0 & 1 & \dots & 0\\
        \vdots & \vdots & \vdots & \ddots & \vdots\\
        0 & 0 & 0 & \dots & 1\\
        a_1 & a_2 & a_3 & \dots & a_n
        \end{bmatrix}
        \quad & & B = \begin{bmatrix}
        0 \\
        0 \\
        \vdots \\
        0\\
        b
        \end{bmatrix}
    \end{aligned}
\end{equation}\label{eq:A.1.2}
Where the values $a_{1, 2 \ldots, n}, b \neq 0$ are unknown and the signs are known. We are given a known reference model in the same controllable canonical form:
\begin{equation}
    \dot{x}_r = A_rx_r + B_ru_r
\end{equation}
The subscript $r$ denotes that the parameters and signals are known and belong to the reference model. For notational simplicity, we denote the last row of the matrix $A_r$ by $\alpha_r = [a_{1, r}, a_{2, r}, \ldots, a_{n, r}]^T$ and the non-zero element of $B_r$ by $b_r$. The goal is to track a system in the form of equation \ref{eq:A.1.1}:
\begin{equation}\label{eq:A.1.3}
    \dot{x} = Ax + \lambda B_r u
\end{equation}
Where we have used knowledge of the form of $B, B_r$ to slightly rewrite the equation. $\lambda > 0$ is an unknown scalar, $A$ is an unknown matrix (in the same form as \eqref{eq:A.1.2}, with signs known) and $B_r$ is known. We would then like to determine an input $u(t)$ to the system in equation \ref{eq:A.1.3} such that $\lim\limits_{t \to \infty}||e(t)|| = 0$, where we have defined $e(t) \coloneqq x(t) - x_r(t)$. If the true system parameters are known, the following ideal control law provides perfect tracking of the reference system:
\begin{equation}
    \begin{aligned}
    u^* = K_x^Tx + k_u u_r
    \end{aligned}
\end{equation}
With $K_x, \; k_u$ satisfying the following matching conditions:
\begin{equation}\label{eq:A.1.4}
\begin{aligned}
A+\lambda B_r K_x^T = A_r \quad \lambda k_u B_r = B_r \rightarrow \lambda k_u = 1
\end{aligned}
\end{equation}
We define the diagonal matrix $D \in \mathbb{R}^{nxn}$, with components defined as:
\begin{equation}
    D_{ij} = \begin{cases} 
      \omega_{i} & \alpha_{i, r} > 0 \quad \textrm{and} \quad i = j\\
      \psi_{i} & \alpha_{i, r} < 0 \quad \textrm{and} \quad i = j\\
      0 & i\neq j\\
   \end{cases}
\end{equation}
$\omega_i, \psi_i$ are user defined constant scalars, such that $\omega_i > 1$, $\psi_i \leq 0$ for $i = 1, 2, \ldots, n$. Note, then, that the vector defined by $v = \alpha_r - D\alpha_r$ contains entirely strictly negative values. Furthermore, we define the vector $h \in \mathbb{R}^n$ as $[0, 0, \ldots, 1]^T$. Then, the matrix $A_H$ defined by:
\begin{equation}\label{eq:A.1.4.5}
    \begin{aligned}
    A_H = A_r - h(D\alpha_r)^T
    \end{aligned}
\end{equation}
Is Hurwitz. Consider the following control law:
\begin{equation}
    u = \hat{K}^T_xx + \hat{k}_u u_r - \frac{1}{b_r}\hat{k}_u (D \alpha_r)^T e
\end{equation}
Where $\hat{K}_x$ and $\hat{k}_u$ represent adaptive estimates of $K_x$, $k_u$ respectively. The error dynamics are then:
\begin{flalign*}
\dot{e} &= \dot{x} - \dot{x}_r &&\\
\dot{e} &= Ax - A_rx_r - B_ru_r &&\\
&+\lambda B_r(\hat{K}^T_xx + \hat{k}_u u_r - \frac{1}{b_r}\hat{k}_u (D \alpha_r)^Te)
\end{flalign*}
Noting that $B_r = [0, 0, \ldots, b_r]^T \Rightarrow \frac{1}{b_r}B_r = h$, we then have:
\begin{flalign*}
\dot{e} &= Ax - A_rx_r - B_ru_r &&\\
&+ \lambda B_r \hat{K}^T_xx + \lambda B_r \hat{k}_u u_r - \lambda \hat{k}_u h (D \alpha_r)^T e
\end{flalign*}
Utilizing the matching conditions \eqref{eq:A.1.4}:
\begin{flalign*}
\dot{e} &= (A_r - \lambda B_r K_x^T)x - A_rx_r - \lambda k_u B_ru_r &&\\
&+ \lambda B_r \hat{K}^T_xx + \lambda B_r \hat{k}_u u_r - \lambda \hat{k}_u h (D \alpha_r)^T e &&\\
\dot{e} &= A_r(x - x_r) - \lambda \hat{k}_u h(D\alpha_r)^Te &&\\
&+ \lambda B_r [(\hat{K}^T_x - K^T_x)x + (\hat{k}_u - k_u)u_r]
\end{flalign*}
Again utilizing the matching condition $\lambda k_u = 1$, and defining the parameter estimation errors $\Tilde{K}_x = \hat{K}_x - K_x$, $\Tilde{k}_u = \hat{k}_u - k_u$:
\begin{flalign*}
\dot{e} &= A_re -h(D\alpha_r)^Te + \lambda k_u h(D\alpha_r)^Te - \lambda \hat{k}_u h(D\alpha_r)^Te + \lambda B_r [\Tilde{K}^T_xx + \Tilde{k}_uu_r] &&\\
\dot{e} &= A_re -h(D\alpha_r)^Te + \lambda B_r [\Tilde{K}^T_xx + \Tilde{k}_u(u_r - \frac{1}{b_r}(D\alpha_r)^Te)]
\end{flalign*}
Applying \eqref{eq:A.1.4.5} and defining an augmented reference input $\xi = u_r - \frac{1}{b_r}(D\alpha_r)^Te$, the error dynamics are then:
\begin{equation}\label{eq:A.1.5}
    \begin{aligned}
        \dot{e} = A_H e + \lambda B_r [\Tilde{K}^T_xx + \Tilde{k}_u\xi]
    \end{aligned}
\end{equation}
\noindent Now consider the Lyapunov function candidate:
\begin{equation}\label{eq:lyapunov}
V(e, \Tilde{K}_x, \tilde{k}_u) = e^TPe + \lambda \Tr(\Tilde{K}^T_x\Gamma_x^{-1}\Tilde{K}_x) + \lambda\frac{\Tilde{k}_u^2}{\gamma_u}
\end{equation}
\noindent With $\Gamma_x$ positive definite and scalar $\gamma_u > 0$. We have also introduced a $P = P^T \succ 0$ that satisfies the Lyapunov equation:
\begin{equation*}
    PA_H + A^T_HP = -Q \quad
    \textrm{with} \quad Q = Q^T \succ 0
\end{equation*}
Because $P \succ 0$ the Lyapunov function $V$ is positive definite. The time derivative is then calculated:
\begin{flalign*}
\dot{V} &= \dot{e}^TPe + e^TP\dot{e} + 2\lambda\Tr(\Tilde{K}^T_x\Gamma_x^{-1}\dot{\hat{{K}}}_x) + 2\lambda\frac{\Tilde{k}_u\dot{\hat{k}}_u}{\gamma_u}&&\\
\dot{V} &=(A_He + \lambda B_r[\Tilde{K}_x^Tx + \Tilde{k}_u\xi ])^TPe + e^TP(A_He + \lambda B_r[\Tilde{K}_x^Tx + \Tilde{k}_u\xi ])&&\\
&+2\lambda\Tr(\Tilde{K}^T_x\Gamma_x^{-1}\dot{\hat{{K}}}_x) + 2\lambda\frac{\Tilde{k}_u\dot{\hat{k}}_u}{\gamma_u}&&\\
\dot{V} &= e^TA^T_HPe + e^TPA_He &&\\
&+ 2\lambda[e^TPB_r\Tilde{K}_x^Tx + \Tr(\Tilde{K}^T_x\Gamma_x^{-1}\dot{\hat{{K}}}_x)]&&\\
&+ 2\lambda [e^TPB_r\Tilde{k}_u\xi  + \frac{\Tilde{k}_u\dot{\hat{k}}_u}{\gamma_u}]&&\\
\dot{V} &= -e^TQe + 2\lambda[e^TPB_r\Tilde{K}_x^Tx + \Tr(\Tilde{K}^T_x\Gamma_x^{-1}\dot{\hat{{K}}}_x)] + 2\lambda [e^TPB_r\Tilde{k}_u\xi  + \frac{\Tilde{k}_u\dot{\hat{k}}_u}{\gamma_u}]
\end{flalign*}
\noindent Note, for column vectors $a, b$, we have $a^Tb = \Tr(ba^T)$. Then, $(e^TPB_r)(\Tilde{K}_x^Tx) = \Tr(\Tilde{K}_x^Txe^TPB_r)$:
\begin{flalign*}
\dot{V} &= -e^TQe + 2\lambda\Tr(\Tilde{K}_x^T[xe^TPB_r + \Gamma_x^{-1}\dot{\hat{{K}}}_x]) + 2\lambda \Tilde{k}_u[e^TPB_r\xi  + \frac{\dot{\hat{k}}_u}{\gamma_u}]&&\\
\end{flalign*}
\noindent By defining the adaptive parameter update laws as:
\begin{equation}\label{eq:Kxd}
    \dot{\hat{K}}_x = -\Gamma_xxe^TPB_r
\end{equation}
\begin{equation}\label{eq:Kxd2}
    \dot{\hat{k}}_u = -\gamma_u\xi e^TPB_r
\end{equation}
\noindent we find that the Lyapunov function time derivative is negative semi-definite:
\begin{equation*}
    \dot{V} = -e^TQe \leq 0
\end{equation*}
\noindent Which implies that the tracking error vector $e(t)$ and the parameter estimation errors are bounded and that $V$ (given by \ref{eq:lyapunov}) is a Lyapunov function.
\end{proof}

\begin{manualtheorem}{\ref{corollary:control_law}}
    The MRAC system in Section \ref{ss: nonlinear model} with control law \ref{eq:lin_cntrl_law} and parameter update laws \ref{eq:lin_param_update} exhibits global uniform asymptotic tracking of the reference model dynamics \ref{eq:lin_ref_mod_cc}, for any bounded reference input $u_r(t)$ that generates bounded signals in the reference model. That is: $\lim\limits_{t \to \infty}||x(t) - x_r(t)|| = 0$ if $||x_r(t)|| < M_x$ and $||u_r(t)|| < M_u \; \forall t \in [0, T]\;$ for $M_x, M_u > 0$
\end{manualtheorem}
\begin{proof}\label{proof1}
It is assumed that the reference input $u_r(t)$ is bounded and results in a reference system with bounded states. Note that this is a condition imposed on the trained reinforcement learning policy $\pi$ - namely that the learned policy generates bounded responses in simulation. From Theorem \ref{theorem:control_law}, the tracking error $e(t)$ is uniformly bounded and stable, and the parameter estimates $\hat{K}_x(t)$ and $\hat{k}_r(t)$ are uniformly bounded. From the assumption, $x_r(t), \dot{x_r}(t)$ are bounded, and thus $x(t) = e(t) + x_r(t)$ is bounded. The boundedness of $\hat{K}_x, \hat{k}_r, x, u_r$ then implies boundedness of $u(t)$, which then implies the boundedness of $\dot{x} = Ax + \lambda B_r u$. Thus $\dot{e} = \dot{x} - \dot{x}_r$ is bounded. A direct result is that the second time derivative of V:
\begin{equation*}
    \ddot{V} = -2e^TQe
\end{equation*}
is bounded. Thus, $\dot{V}$ is uniformly continuous. Because $V(t) \geq 0$ and $\dot{V}(t) \leq 0$ we have from Barbalat's Lemma that $\lim\limits_{t \to \infty}\dot{V}(t) = 0$. Hence, $\lim\limits_{t \to \infty}||e(t)|| = 0$: the tracking error tends to the origin globally, uniformly and asymptotically. 
\end{proof}

\subsection{Adaptive Controller: Nonlinear System}

\begin{manualtheorem}{\ref{theorem:control_law_nl} (Summary)}
    Using the following control law and adaptive parameter update laws:
    \begin{equation}
        \begin{aligned}
            u = \hat{K}_\zeta^T\zeta + \hat{k}_uu_r - \frac{1}{b_r}\hat{k}_u \alpha_r^T(\zeta - \zeta_r) + \frac{1}{b_r}\hat{k}_u \beta_r^T e
        \end{aligned}
    \tag{\ref{eq:nlin_cntrl_law}}
    \end{equation}
    \begin{equation}
        \begin{aligned}
            \dot{\hat{K}}_\zeta = -\Gamma_\zeta\zeta ^TPB_r, \quad \dot{\hat{k}}_u = -\gamma_u\xi e^TPB_r
        \end{aligned}
    \tag{\ref{eq:nlin_param_update}}
    \end{equation} 
$V(e, \Tilde{K}_x, \tilde{k}_u) = e^TPe + \lambda \Tr(\Tilde{K}^T_x\Gamma_x^{-1}\Tilde{K}_x) + \lambda\frac{\Tilde{k}_u^2}{\gamma_u}$ is a Lyapunov function
\end{manualtheorem}

\begin{proof}
We proceed in a manner similar to the proof in \ref{ss: linear}. Nonlinear dynamical systems of the following form are considered:
\begin{equation}\label{eq:A.2.1}
    \begin{aligned}
    \dot{x} = A\zeta (x) + B u
    \end{aligned}
\end{equation}
Where $\zeta : \mathbb{R}^n \to \mathbb{R}^n$ is a known nonlinear map of the form:
\begin{equation*}
    \begin{aligned}
    \zeta (x) = [\phi(x_1), x_2, x_3, \ldots, x_n]^T
    \end{aligned}
\end{equation*}
With $\phi: \mathbb{R} \to \mathbb{R}$ a known nonlinear function. For notational simplicity, we use $\zeta$ to refer to $\zeta (x)$. In equation \ref{eq:A.2.1}, matrix $A$ is unknown (but the signs of the entries are known) and matrix $B$ is unknown. Furthermore, the pair $(A, B)$ are in a "pseudo"-controllable canonical form. That is, $A$ and $B$ are of the form:
\begin{equation}\label{eq:A.2.2}
    \begin{aligned}
        A = \begin{bmatrix}
        0 & 1 & 0 & \dots & 0\\
        0 & 0 & 1 & \dots & 0\\
        \vdots & \vdots & \vdots & \ddots & \vdots\\
        0 & 0 & 0 & \dots & 1\\
        a_1 & a_2 & a_3 & \dots & a_n
        \end{bmatrix}
        \quad & & B = \begin{bmatrix}
        0 \\
        0 \\
        \vdots \\
        0\\
        b
        \end{bmatrix}
    \end{aligned}
\end{equation}
Where the values $a_{1, 2 \ldots, n}, b$ are unknown and the signs are known. We are given a known reference model in the same form as equation \ref{eq:A.2.1}:
\begin{equation}
    \begin{aligned}
    \dot{x}_r = A_r\zeta_r + B_r u_r
    \end{aligned}
\end{equation}
With corresponding reference signals $x_r, \zeta_r, u_r$. $(A_r, B_r)$ are in the same "pseudo"-controllable canonical form:
\begin{equation}\label{eq:A.2.3}
    \begin{aligned}
        A_r = \begin{bmatrix}
        0 & 1 & 0 & \dots & 0\\
        \vdots & \vdots & \vdots & \ddots & \vdots\\
        a_{1, r} & a_{2, r} & a_{3, r} & \dots & a_{n, r}
        \end{bmatrix}
        \quad & & B_r = \begin{bmatrix}
        0 \\
        \vdots \\
        b_r
        \end{bmatrix}
    \end{aligned}
\end{equation}
Where the subscript r denotes that the parameters are known and belong to the reference model. For notational simplicity, we use the vector $\alpha_r = [a_{1, r}, a_{2, r}, \ldots, a_{n, r}]^T$ to compactly represent the last row of $A_r.$ The goal is to track a system with dynamics of the form:
\begin{equation}\label{eq:A.2.4}
    \begin{aligned}
    \dot{x} = A\zeta + \lambda B_r u
    \end{aligned}
\end{equation}
Note, that this model form is equivalent to the form presented in equation \ref{eq:A.2.1}. We have just used the known structures of $B, B_r$ and the introduction of an unknown scalar $\lambda > 0$ to slightly rewrite the system. We would then like to choose an input to the system in equation \ref{eq:A.2.4} such that $\lim\limits_{t \to \infty}||e(t)|| = 0$, where we have defined $e(t) \coloneqq x(t) - x_r(t)$ 
\noindent If the true system parameters are known, the following ideal control law provides perfect tracking of the reference system: 
\begin{equation}\label{eq:ideal_law_nl}
    u^* = K_\zeta^T\zeta + k_uu_r
\end{equation}
With $K_\zeta, \; k_u$ satisfying the matching conditions:
\begin{equation}\label{eq:match_conditions_ap_nl}
\begin{aligned}
A+\lambda B_r K_\zeta^T = A_r \quad \lambda k_u B_r = B_r \rightarrow \lambda k_u = 1
\end{aligned}
\end{equation}
Consider the following adaptive control law:
\begin{equation}\label{eq:general_ctrl}
    u = \hat{K}_\zeta^T\zeta + \hat{k}_uu_r - \frac{1}{b_r}\hat{k}_u \alpha_r^T(\zeta - \zeta_r) + \frac{1}{b_r}\hat{k}_u \beta_r^T e
\end{equation}
Where $\hat{K}_\zeta$ and $\hat{k}_u$ represent adaptive estimates of $K_\zeta$, $k_u$ respectively. The error dynamics are then:
\begin{flalign*}
\dot{e} &= \dot{x} - \dot{x}_r &&\\
\dot{e} &= A\zeta - A_r \zeta_r - B_r u_r &&\\
&+ \lambda B_r [\hat{K}_\zeta^T\zeta + \hat{k}_uu_r - \frac{1}{b_r}\hat{k}_u \alpha_r^T(\zeta - \zeta_r) + \frac{1}{b_r}\hat{k}_u \beta_r^T e]
\end{flalign*}
Defining the the vector $h \in \mathbb{R}^n$ as $[0, 0, \ldots, 1]^T$ and noting that $B_r = [0, 0, \ldots, b_r]^T \Rightarrow \frac{1}{b_r}B_r = h$, we then have:
\begin{flalign*}
\dot{e} &= A\zeta - A_r \zeta_r - B_r u_r &&\\
&+ \lambda B_r \hat{K}_\zeta^T\zeta + \lambda B_r \hat{k}_uu_r - \lambda \hat{k}_u h \alpha_r^T(\zeta - \zeta_r) + \lambda \hat{k}_u h \beta_r^T e
\end{flalign*}
We now utilize the matching conditions  \eqref{eq:match_conditions_ap_nl}:
\begin{flalign*}
\dot{e} &= (A_r - \lambda B_r K_\zeta^T)\zeta - A_r \zeta_r - \lambda k_u B_r u_r &&\\
&+ \lambda B_r \hat{K}_\zeta^T\zeta + \lambda B_r \hat{k}_uu_r - \lambda \hat{k}_u h \alpha_r^T(\zeta - \zeta_r) + \lambda \hat{k}_u h \beta_r^T e &&\\
\dot{e} &= A_r(\zeta - \zeta_r) + \lambda B_r \zeta (\hat{K}_\zeta^T - K_\zeta^T) &&\\
&+ \lambda B_r u_r (\hat{k}_u - k_u) - \lambda \hat{k}_u h \alpha_r^T(\zeta - \zeta_r) + \lambda \hat{k}_u h \beta_r^T e &&\\
\end{flalign*}
Recall, $\zeta = [\phi(x_1), x_2, \ldots, x_n]^T$. Then, $\zeta - \zeta_r = [\phi(x_1) - \phi(x_{1_r}), e_2, e_3, \ldots e_n]^T$. Additionally, note that $A_r$ may be rewritten in the following form:
\begin{equation}
\begin{aligned}
A_r = \left[
\begin{array}{c}
    \begin{array}{c | c}
        \vec{O}_{(n-1)\times 1} & \vec{I}_{(n-1)\times(n-1)}
    \end{array} \\
\hline
\alpha_r^T\\
\end{array}
\right] = M + h\alpha_r^T
\end{aligned}
\end{equation}
Where we have defined $M \coloneqq [\begin{array}{c}
    \begin{array}{c | c}
        \vec{O}_{(n-1)\times 1} & \vec{I}_{(n-1)\times(n-1)}
    \end{array} \\
\hline
\vec{O}_{1\times n}\\
\end{array}]$. We may then equivalently write:\\\\
$A_r(\zeta - \zeta_r) =  Me + h\alpha_r^T(\zeta - \zeta_r)$\\\\
We utilize this substitution, along with the matching condition $\lambda k_u = 1$, to write the error dynamics as:
\begin{flalign*}
\dot{e} &= Me + \lambda k_u h\alpha_r^T(\zeta - \zeta_r) - \lambda \hat{k}_u h \alpha_r^T (\zeta - \zeta_r) &&\\
&+ h \beta_r^T e - \lambda k_u h\beta_r^T e + \lambda \hat{k}_u h \beta_r^T e &&\\
&+ \lambda B_r \zeta (\hat{K}_\zeta^T - K_\zeta^T) + \lambda B_r u_r(\hat{k}_u - k_u)
\end{flalign*}
Defining the parameter estimation errors, $\Tilde{K}_\zeta = \hat{K}_\zeta - K_\zeta$ and $\Tilde{k}_u = \hat{k}_u - k_u$, and noting that $M+h\beta_r^T = A_H$, we have:
\begin{flalign*}
\dot{e} &= A_He + \lambda B_r [\Tilde{K}_\zeta^T \zeta + \Tilde{k}_u(u_r - \frac{1}{b_r}\alpha_r^T(\zeta - \zeta_r) + \frac{1}{b_r}\beta_r^Te)] &&\\
\end{flalign*}
Defining the augmented reference input $\xi (t) \coloneqq u_r(t) - \frac{1}{b_r}\alpha_r^T(\zeta(t) - \zeta_r(t)) + \frac{1}{b_r}\beta_r^Te(t)$, the error dynamics are then compactly represented as:
\begin{equation}
    \begin{aligned}
    \dot{e} = A_He + \lambda B_r[\Tilde{K}_\zeta^T \zeta + \Tilde{k}_u \xi]
    \end{aligned}
\end{equation}
\noindent Now we prove that $\lim\limits_{t \to \infty}||e(t)|| = 0$. Construct the following Lyapunov function:
\begin{equation*}
V(e, \Tilde{K}_\zeta, \tilde{k}_u) = e^TPe + \lambda \Tr(\Tilde{K}^T_\zeta\Gamma_\zeta^{-1}\Tilde{K}_\zeta) + \lambda\frac{\Tilde{k}_u^2}{\gamma_u}
\end{equation*}
By the same exact procedure described in \ref{ss: linear}, we find that $\dot{V} = -e^TQe \leq 0$ when the following adaptive control laws are defined:
\begin{equation*}
    \dot{\hat{K}}_\zeta = -\Gamma_\zeta\zeta e^TPB_r
\end{equation*}
\begin{equation*}
    \dot{\hat{k}}_u = -\gamma_u\xi e^TPB_r
\end{equation*}
\noindent Thus V is a Lyapunov function and tracking error and parameter estimation errors are bounded.
\end{proof}

\begin{manualtheorem}{\ref{corollary:nl_control_law}}
\end{manualtheorem}
\begin{proof} The proof follows from the proof of Theorem \ref{corollary:control_law}
\end{proof}

\section{Simulation and Training Details}\label{app:b}
The following equations are used to simulate the inverted pendulum:
\begin{equation}
    \begin{aligned}
         ml^2\ddot{\theta} = mgl\theta -b\dot{\theta} + u\quad
        \textrm{(linear)} \quad \quad ml^2\ddot{\theta} = mgl\sin\theta -b\dot{\theta} + u \quad \textrm{(nonlinear)}
    \end{aligned}
\end{equation}
We use the following (unitless) nominal parameter values for the reference model: $m = 1, \; l =1, \; b = 1, \; g = 10$. In order to simulate the reference model, we utilize Euler's method with a numerical integration frequency of $200Hz$.

Given the dynamics model, we then define the objective. For SRIP, a desired angular set-point, $\theta_0(t)$, is provided, and changes randomly every 5 seconds. The optimal control objective is then to minimize the following expression:  
\begin{equation}
    \begin{aligned}
    \sum_{k=0}^{T}q_1(\theta(k) - \theta_0)^2 + q_2\dot{\theta}(k)^2 + ru(k)^2
    \end{aligned}
\end{equation}

\noindent which is a typical quadratic cost. We use an episode length of 20 seconds and an agent interaction frequency of $10Hz$. As a result, the number of episode steps is given by $T = 200$. In our implementation we set $q_1 = 1.0, \; q_2 = .1 ,\; r = .001$.

\begin{table}[H]
  \begin{center}
    \begin{tabular}{c|c  c|c  c}
      \toprule 
      \textbf{Algorithm} & \multicolumn{2}{c|}{\textbf{Linear Model}} & \multicolumn{2}{c}{\textbf{Nonlinear Model}}\\
      \toprule
       & \textbf{Average Cost} & 
      \textbf{Hyperparameters} & \textbf{Average Cost} & 
      \textbf{Hyperparameters}\\

      \midrule 
      PPO & 103 & \begin{tabular}{@{}c@{}}[$\gamma: .99, \;$\\ lr*$: 8\mathrm{e-}5$, \\ ent\_coeff$: 0.001$,\\ total\_timesteps$: 2\mathrm{e}5$]\end{tabular}
      
      & 177 & \begin{tabular}{@{}c@{}}[$\gamma: .99, \;$ \\ lr*$: 2\mathrm{-e}5$, \\ ent\_coeff$: 0.001$,\\ total\_timesteps$: 9\mathrm{e}5$]\end{tabular}\\
      \midrule
            DDPG & 132 & \begin{tabular}{@{}c@{}}[$\gamma: .99, \;$\\ lr*$: 7\mathrm{e-}4$, \\ Noise: OU**,\\ total\_timesteps$: 7\mathrm{e}5$]\end{tabular}
      
      & 264 & \begin{tabular}{@{}c@{}}[$\gamma: .99, \;$\\ lr*$: 7\mathrm{e-}4$, \\ Noise: OU**,\\ total\_timesteps$: 8\mathrm{e}6$]\end{tabular}\\
      
    \midrule
            SAC & 78  & \begin{tabular}{@{}c@{}}[$\gamma: .99, \;$\\ lr*$: 5\mathrm{e-}4$, \\ total\_timesteps$: 8\mathrm{e}4$]\end{tabular}
      
      & 151  & \begin{tabular}{@{}c@{}}[$\gamma: .99, \;$\\ lr*$: 5\mathrm{e-}4$, \\ total\_timesteps$: 9\mathrm{e}5$]\end{tabular}\\
      
      \bottomrule 
    \end{tabular}
      \caption{Reinforcement learning training details. Average cost is measured as the average return over 100 episodes. For each algorithm, the most salient hyperparameter values are provided. Hyperparameters were chosen via simple grid search. The RL agents are trained using a 10Hz environment interaction frequency.\\
      * Learning rates are for both actor and critic networks\\
      ** Ornstein-Uhlenbeck process with $\mu = 0$, $\sigma = 1.5$}
      \label{tab:table2}
  
  \end{center}

\end{table}
\noindent We then utilize three popular reinforcement learning algorithms (PPO, SAC, DDPG) to train control policies for this environment. We utilize the Stable Baselines (\citealt{stable-baselines}) implementations of these algorithms. Stable Baselines provides a number of high quality RL algorithms, and is based on the popular OpenAI Baselines implementations. Training details are provided in Table \ref{tab:table2}.

After training on the reference models, we create $1000$ "test" environments for each of the linear and nonlinear pendulum models. For each test environment, model parameters are randomly sampled from the following ranges: $l \in [.75, 1.25], \; m \in [.75, 1.25], \; b \in [0.001, 2.0]$. Each test environment is also associated with a sequence of four angular set-points, sampled as: $\theta_0^i \in [-\pi, \pi], \; i = 1, 2, 3, 4$. We then evaluate the performance of the various inner-outer loop algorithms on these test environments. The use of "RL" in Table \ref{tab:table1} indicates the aggregation of results from using PPO, DDPG and SAC. For example, values provided for \textbf{[10Hz RL; 100Hz MRAC]} are calculated as the average of the values from \textbf{[10Hz PPO; 100Hz MRAC]}, \textbf{[10Hz DDPG; 100Hz MRAC]} and \textbf{[10Hz SAC; 100Hz MRAC]}

\end{document}